\documentclass[12pt]{article}
\usepackage{amsfonts,amsmath,amssymb,amscd,mathrsfs,enumitem,graphicx}
\usepackage{breakcites}
\usepackage{natbib}
\usepackage{braket,url,hyperref}
\usepackage{pifont}

\title{On substantive Lorentz invariance and quantum theory}

\author{Ward Struyve\footnote{Department of Physics and Astronomy, KU Leuven, Belgium}$^{*}$\footnote{Centre for Logic and Philosophy of Science, KU Leuven, Belgium}  }

\date{}

\def\si{\sigma}
\def\Si{\Sigma}

\def\lam{\lambda}

\def\pa{\partial}

\def\ii{\textrm i}
\def\ee{\textrm e}
\def\e{\textrm e}

\def\ud{\textrm{d}}

\newcommand{\be}{\begin{equation}}
\newcommand{\en}{\end{equation}}
\newcommand{\bi}{\begin{itemize}}
\newcommand{\ei}{\end{itemize}}

\begin{document}
\maketitle

\begin{abstract}
	\noindent
Lorentz invariance is considered a fundamental property of relativistic quantum theory. However, in standard quantum theory this invariance is only partially realized: while relativistic wave equations like the Dirac equation are Lorentz invariant, the collapse postulate is not. In the non-relativistic domain, alternative theories have been formulated, like Bohmian mechanics and spontaneous collapse theories, which dispense with the problematic collapse postulate. Extending these theories to the relativistic domain appears challenging, particularly with respect to implementing Lorentz invariance, mainly due to the unavoidable non-locality implied by Bell's theorem. However, space-time theories can also trivially be formulated in a Lorentz-invariant way. If Lorentz invariance is to impose a genuine constraint on the content of a physical theory, one should aim for what can be called {\em substantive Lorentz invariance}. However, articulating a precise definition of this notion is notoriously difficult. This paper investigates two candidate criteria: Anderson’s criterion based on the identification of absolute objects, and a relativity principle for isolated subsystems. These criteria are applied to evaluate several Lorentz-invariant Bohmian models, a spontaneous collapse model, and a version of the Many-Worlds theory. With the exception of two Bohmian approaches, these models satisfy both criteria. Nevertheless, some Bohmian models that meet these conditions still do not appear to be substantively Lorentz invariant, suggesting that the proposed criteria may not fully capture the intended concept. Nevertheless, they help clarify what aspects of relativity theory may need to be given up in passing from classical to quantum theory.
\end{abstract}

\section{Introduction}
Bohmian mechanics \citep{holland93b,bohm93,duerr09} and spontaneous collapse theories \citep{bassi03} provide alternatives to standard quantum theory that solve the measurement problem. Although these theories are fully developed and well understood in the non-relativistic domain, the construction of relativistic extensions is considerably more challenging. One challenge concerns the incorporation of particle creation and annihilation. This issue has been considered in the Bohmian context \citep{duerr04b,duerr05a,colin07,struyve10,struyve11a}, as well as in spontaneous collapse models \citep{tumulka06d}. Another challenge, which is the main concern of this paper, is the implementation of Lorentz invariance. In their simplest formulation, these models employ a preferred Lorentz frame.
Although the frame may be empirically inaccessible, so that statistical predictions remain Lorentz invariant, the fact that the dynamics depends on a specific frame implies a violation of Lorentz invariance at the fundamental level.
Only in recent years, some (toy) models that are fully Lorentz invariant have been proposed~\citep{berndl96a,duerr99,horton01,dewdney02,goldstein03,horton04,nikolic05b,tumulka07,duerr14a,tumulka14,galvan15,tumulka06a,tumulka21,bedingham11,bedingham14,pearle15,tumulka22}. 

Achieving Lorentz invariance is particularly difficult because these theories must be non-local, i.e., they necessarily involve influences over space-like distances, as implied by Bell's theorem \citep{bell87a,maudlin02,goldstein11b}. Non-locality, however, does not preclude Lorentz invariance. Lorentz invariance concerns the symmetry properties of a theory under space-time transformations and a theory may well be non-local while still respecting these symmetries. Familiar examples include tachyonic theories and the Wheeler-Feynman absorber theory of electrodynamics. Nevertheless, in the context of quantum mechanics, it seems difficult to reconcile the non-locality with Lorentz invariance. 

At first sight, it might sound strange that Lorentz invariance is hard to achieve. After all, every textbook on relativistic quantum mechanics or quantum field theory is concerned with such formulations. However, these textbooks tend to ignore the collapse postulate. According to standard quantum mechanics, the quantum state vector is supposed to collapse upon measurement. However, if the collapse is instantaneous, it requires the specification of a particular time, which directly conflicts with Lorentz invariance. 
 
Even setting aside the issue of collapse, there is another source of tension with Lorentz invariance. For a theory defined on space-time, it is natural to demand that the symmetries of the space-time also yield symmetries of the dynamical laws~\citep[p.\ 46]{earman89}. So for a theory formulated on special relativistic space-time, i.e., Minkowski space-time, it is natural to demand that the laws of motion be Lorentz invariant. However, standard quantum theory does not appear to be a theory that is defined on space-time. Its primary ontological candidate is the quantum state vector, but that is not an object on physical space. Instead, it tends to be understood as an object on configuration space, or more abstractly as an element of some Hilbert space. This feature not only raises the difficulty of accounting for the appearance of entities in space-time and space-time itself \citep{norsen17a,maudlin19b}, but also obscures the possible connection between Lorentz invariance and space-time symmetries. If the fundamental ontology of the theory is not spacetime-based, the requirement that the state vector transforms according to a representation of the Lorentz group becomes dubious.

The latter issue is absent in Bohmian and spontaneous collapse theories. These theories concern an ontology that is defined on space-time, sometimes referred to as the {\it primitive ontology} \citep{allori08}. In non-relativistic Bohmian mechanics the primitive ontology consists of point-particles. In the Ghirardi, Rimini and Weber (GRW) collapse theory, two possible choices are the mass density ontology and the flash ontology, resulting in the GRWm and GRWf theories. The wave function, which still plays a role in all these theories, may be viewed as part of the ontology, but it is not part of the primitive ontology. (Instead, it can also be viewed as part of the law governing the dynamics of the primitive ontology \citep{goldstein13}.)

The Many-Worlds theory is another prominent solution to the measurement problem~\citep{barrett99,saunders10,wallace12}. In its standard formulation, the ontology consists solely of the universal quantum state, which evolves unitarily at all times; no additional variables are introduced to provide an ontology in space-time. In a relativistic setting, the theory is fully Lorentz invariant. However, in the absence of a primitive ontology, the challenge remains of explaining how the familiar macroscopic world emerges and to connect the Lorentz invariance to space-time symmetries. Recent proposals have sought to supplement the theory with an explicit space-time ontology.  One such proposal, the Sm theory, follows a strategy similar to that of the GRWm theory by introducing a mass density in the non-relativistic case and an energy-momentum tensor in the relativistic setting~\citep{allori11}. 
Another proposal assigns a state associated to each space-time region, based on the algebraic quantum field theory notion of state \citep{wallace10}. Although these proposals help to account for the macroscopic world, challenges remain in making the account precise~\citep{allori11,maudlin19b}. Nonetheless, among the challenges these theories may face, Lorentz invariance is not one of them. In Bohmian mechanics and spontaneous collapse theories, the difficulty lies in formulating a Lorentz-invariant dynamics for the primitive ontology. By contrast, in the Many-Worlds approaches \citep{allori11,wallace10}, the primitive ontology is derived from the quantum state and thus inherits the Lorentz invariance of its unitary evolution. (Whether these Many-Worlds theories are non-local is a more subtle issue. Bell's theorem assumes single outcomes of measurements, which is not the case in the Many-Worlds approach, and hence has no bearing on the issue.)

As we shall detail in the next section, there are also trivial ways of rendering space-time theories Lorentz invariant. For example, even Newtonian mechanics can be reformulated in a Lorentz-invariant manner. If the requirement of Lorentz invariance is to impose any constraint on theory construction, a distinction is needed between what might be called {\em substantive} Lorentz invariance and mere {\em formal} Lorentz invariance. However, the difficulty lies in making this distinction precise. Bell was particularly concerned with this issue, but did not see a way to define what he termed {\em serious Lorentz invariance} \citep{bell84}. Apart from Bell, this problem has received little attention in the context of quantum mechanics. 

On the other hand, there has been an extensive debate on a related issue in general relativity  \citep{anderson67,anderson71,friedman83,norton93,earman06a,earman06b,giulini07,pitts06,pooley10,pooley17,read24}.\footnote{This debate is still ongoing. Norton’s subtitle “Eight Decades of Dispute” aptly summarizes the situation \citep{norton93}.} 
While Einstein developed his theory by requiring general covariance, it was argued by \citet{kretschmann17} that this motivation is physically vacuous, since any space-time theory can trivially be reformulated in a generally covariant way. In response, several proposals have been put forward to articulate a notion of substantive general covariance. The aim of this paper is to apply two of the more promising approaches to the issue of Lorentz invariance. The first is due to \citet{anderson67,anderson71} and is based on the distinction between absolute and dynamical objects. The second appeals to a relativity principle for (approximately) isolated subsystems~\citep{brownhr95,budden97,healey07,greaves14}.

We will examine several Lorentz-invariant Bohmian models, a relativistic generalization of the GRW theory, called the rGRWf model, and the relativistic version of the Sm theory, assessing them in light of the two characterizations of substantive Lorentz invariance.{\footnote{Apart from some Bohmian models that are novel, these theories are discussed in detail by \citet{tumulka22}.}} 
The discussion is restricted to the case of a fixed number of non-interacting Dirac particles. This assumption keeps the analysis tractable, avoiding the additional complexities of quantum field theory. Although the particles do not interact, the wave function may be entangled, which is the source of non-locality. This analysis not only serves to evaluate the Lorentz invariance in these models, but also allows us to explore the adequacy of these notions of substantive Lorentz invariance beyond the well-trodden domain of classical theories.

We focus here on the possible meaning of Lorentz invariance and to what extent they are satisfied in versions of quantum theory. But this discussion should be viewed in the broader context of what it could mean for a theory to be fundamentally relativistic \citep{duerr14a,allori22,tumulka22}. There are several features one might reasonably demand of a fundamentally relativistic theory. But these seem hard to realize simultaneously in a quantum theory. We will return to this question in the conclusion.

The outline of the paper is as follows. In the next section, we examine different possible notions of substantive Lorentz invariance. Before considering Lorentz-invariant quantum models, it is instructive to consider their non-relativistic counterparts, where no difficulties arise concerning Galilean invariance. This will be done in Section \ref{ch2}, where we discuss Bohmian mechanics, the GRW theory and the Sm version of the Many-Worlds theory in the Galilean context. Section~\ref{ch2-rbm} then turns to relativistic Bohmian mechanics, presenting six distinct Lorentz-invariant formulations. In Section~\ref{ch2-sct}, we consider the rGRWf theory, followed by a brief discussion of the Sm theory in Section~\ref{ch2-mw}. We conclude in Section~\ref{ch2-con}.

\section{Substantive Lorentz invariance}\label{sli}
Let us begin by recalling what it means for a theory to be invariant under a group of space-time symmetries. The discussion will only cover the very basics; for more detailed treatments, the reader is referred to \citep{anderson67,friedman83,earman89}.

A physical theory specifies certain geometric objects defined on a space-time manifold. Some of these objects represent the structure of space-time itself---for example, the Minkowski metric in special relativity---while others represent non-space-time entities, such as particle worldlines or fields. The collection of all geometric objects that are allowed in principle constitutes the set of kinematically possible histories, denoted ${\mathcal K}$. The dynamical laws of the theory then select a subset ${\mathcal D} \subset {\mathcal K}$, consisting of those histories that are dynamically possible. A {\em space-time symmetry} is a map from space-time to itself, with an induced transformation of the geometric objects, such that the objects representing space-time structure are left invariant. The theory is invariant under the space-time symmetry if the symmetry leaves the set ${\mathcal D}$ invariant. In other words, the space-time symmetry maps dynamically possible histories to dynamically possible histories.

In the case of a stochastic theory, invariance requires more \citep{allori08}. Such a theory determines not only which histories are dynamically allowed, but also which probability distributions over those histories are allowed. The theory is invariant under a transformation if that transformation maps any allowed probability distribution to another allowed one. Concretely, given an allowed probability distribution $P(h)$ over histories $h$, the transformed distribution
$P'$ is defined by $P'(h') = P(h)$, where $h'$ is the image of $h$ under the transformation. The transformation qualifies as a symmetry if the resulting distribution $P'$ is itself allowed by the theory.

On this definition, there is a trivial way to reformulate space-time theories that are not Lorentz invariant in a Lorentz-invariant way \cite[p.\ 231]{arntzenius90}, \cite[p.\ 2071]{berndl96a}, \cite[p.\ 376]{tumulka22} (and even in a generally invariant way \citep{kretschmann17,anderson67,friedman83}). To see this, consider a theory that is not invariant and suppose that there is a natural action of the Lorentz group on the geometric objects posited by the theory. The non-invariance implies that there exists a Lorentz transformation $\Lambda$ such that $\Lambda {\mathcal D} \neq {\mathcal D}$. Consider now the theory whose set of dynamically allowed histories is ${\mathcal D}' =  \bigcup_\Lambda  \Lambda {\mathcal D}$, where the union ranges over all possible Lorentz transformations. This is arguably a mere reformulation of the original theory, which is clearly Lorentz invariant, since for any Lorentz transformation~$\Lambda$, we have $\Lambda{\mathcal D}'= {\mathcal D}'$. 

In Newtonian mechanics, the natural space-time structure is that of Galilean space-time, characterized by a spatial metric, a temporal metric, and an affine connection \citep{anderson67,friedman83,earman89}. The material content of the theory is represented by particle worldlines. The theory is invariant under Galilean transformations, but not under Lorentz transformations. Nevertheless, it can trivially be made Lorentz invariant by applying the procedure described above. To perform a Lorentz transformation of the geometric objects, one simply represents them in a Galilean frame and then applies the standard Lorentz transformation. The resulting formulation is Lorentz invariant, but Newton’s laws still hold in a particular Lorentz frame. Hence, aside from Lorentz invariance, this formulation exhibits none of the other characteristic features of a relativistic theory, such as locality, length contraction, etc.

\citet{bell84} (crediting Baumann) suggested a different way for making Newtonian mechanics Lorentz invariant,  by stipulating that Newton's equations hold in the center-of-mass frame determined by the matter distribution (assuming the latter is finite). However, it is not clear how this proposal is supposed to work.{\footnote{ \citet{arntzenius94} raises a similar concern.}} Since the goal is to achieve Lorentz invariance, the center-of-mass frame should be a relativistic one, where the particle worldlines determine the frame in a Lorentz-invariant way. But if Newton's equations must eventually hold, then the worldlines should also admit a Newtonian center-of-mass frame. It is unclear what relativistic notion of center-of-mass frame could guarantee that. (It is even problematic to find a relativistic notion of center-of-mass frame that shares the desired properties of the Newtonian notion \citep{pryce48}.) To illustrate this point, suppose the relativistic center-of-mass frame is defined as a frame where the total relativistic 3-momentum of the particles vanishes (i.e., where the total 4-momentum takes the form $(P^0, 0,0,0)$). However, the problem is that the vanishing of the relativistic 3-momentum does not entail the vanishing of the Newtonian 3-momentum, which is required for a Newtonian center-of-mass frame.

\citet{anderson67,anderson71} refined the notion of what it means for a theory to be invariant under space-time symmetries, see also \citep{friedman73,friedman83,pitts06,giulini07,read24}. Anderson calls the group of transformations that leaves ${\mathcal D}$ invariant, the {\em covariance group} rather than the {\em symmetry group} of the theory. The symmetry group is a subgroup of the covariance group and consists of those transformations that---in our terminology---qualify as substantive symmetry transformations. To define the symmetry group, consider the equivalence relation defined by the group action. An equivalence class under this equivalence relation contains histories that are related by elements of the covariance group. The symmetry group is the largest subgroup of the covariance group that leaves invariant the {\em absolute objects} of the theory. An absolute object is distinguished from a {\em dynamical object}. It is a geometric object that appears in every equivalence class, up to a transformation of the covariance group. In other words, an absolute object is the same for any dynamically possible history, modulo transformations belonging to the covariance group.

What does Anderson's analysis entail for the Lorentz-invariant formulation of Newtonian mechanics? The spatial and temporal metrics of Galilean space-time are clearly absolute objects. But the Lorentz group does not leave these objects invariant and hence is not the symmetry group. In our terminology: the Lorentz invariance is not substantive. 

While Anderson's analysis leads to the desired conclusion in this case, it is not always that successful. Here is an example by \citep{torretti84}. In Newton's theory, the spatial metric is the Euclidean one and qualifies as an absolute object. One could also consider a modified Newtonian dynamics where the spatial metric is non-flat, but still has a constant curvature. Again, this constant curvature metric is an absolute object. But consider now the theory where the metric just has constant curvature, but where the value of the curvature is arbitrary. In the latter theory, the metric no longer is  an absolute object. This result seems counterintuitive, since the metric continues to play the same structural role across those theories. Hence, an absolute object in the theory can be trivially converted into a non-absolute object simply by enlarging the space of dynamically allowed histories. Although the resulting theory is not equivalent to the original theory, it will be empirically adequate wherever the original theory is. Applied to the Lorentz-invariant formulation of Newtonian mechanics, allowing for  spatial metrics with different constant curvature implies that the symmetry group, in Anderson’s sense, would again include the Lorentz group. Other examples of this kind have been considered by \citet{pitts06}.

Another possible way to characterize substantive Lorentz invariance is in terms of a relativity principle for (approximately) isolated subsystems \citep{brownhr95,budden97,healey07,greaves14}. This is inspired by the following observation. Empirical support for dynamical symmetries is typically obtained not by transforming the universe as a whole, but by considering transformations of dynamically isolated subsystems---systems whose evolution is effectively independent of their environment. A symmetry transformation of the whole universe does not produce an observable difference. By contrast, transforming only an isolated subsystem, while leaving its environment unchanged, may yield a physically distinct situation, even though this difference remains undetectable from within the subsystem itself. The prime example is that of Galileo's ship experiment, which can be taken (when suitably qualified) as empirical evidence for invariance under Galilean boosts. The scenarios in which the ship is at rest relative to the shore and in which it moves uniformly relative to it are physically distinct states of affairs, yet they are observationally indistinguishable from within the ship.

A theory can now be said to satisfy a relativity principle for isolated subsystems when solutions of the equations of motion can be transformed into other (approximate) solutions by applying the transformations (Lorentz transformations in the present case) to subsystems that are (approximately) dynamically isolated. For a stochastic theory, also the probabilities for the subsystem should be transformed and match the probabilities determined by the theory for the whole universe. 

Note that we do not require the subsystem to be completely isolated. In practice, this will rarely be the case, since some interaction with the environment typically remains. It suffices, however, that such interactions can be made arbitrarily small in principle. To avoid trivializing this relativity principle, it must also be understood that the theory should allow for approximately isolated subsystems in the first place, and that these subsystems may be of arbitrary complexity. In particular, if the dynamical laws permit a given type of system, they should also permit the existence of multiple such systems within the same universe, perhaps widely separated in space, that evolve approximately independently.

This relativity principle is admittedly somewhat imprecise, but this will suffice for our purposes. Note that in our characterization of it, we did not refer to empirical aspects, as is sometimes done (see, e.g., \citep{budden97}). Our formulation seems more up to the task of investigating substantive Lorentz invariance in the context of quantum mechanics since the latter's statistical character might entail that objects violating Lorentz invariance, like a preferred frame, are unobservable. 

Let us apply this principle to the Lorentz-invariant formulation of Newtonian mechanics. An approximately isolated system will consist of particles that do not (or only very weakly) interact with their environment, e.g., because of substantial spatial separation. But a Lorentz boost of just the system will not lead to a new solution of the dynamical equations (whereas a Galilean boost would). Hence, the Lorentz-invariant formulation of Newtonian mechanics does not satisfy the relativity principle for isolated subsystems with respect to Lorentz transformations.

This should be contrasted with, for example, the Wheeler--Feynman theory, which qualifies as Lorentz invariant in Anderson's sense and also satisfies the relativity principle for subsystems (with a subsystem being dynamically isolated when it is sufficiently distant from its environment).

Finally, one might ask whether the two criteria are related. This question has been investigated by \citet{budden97} and \citet{skow08}, who employ closely related formulations of the criteria but arrive at different conclusions. As we will see in Section \ref{ch2-rbm}, one Bohmian model satisfies Anderson’s criterion while failing to satisfy the relativity principle.

\section{Galilean-invariant non-relativistic quantum mechanics}\label{ch2}
While formulating fully Lorentz-invariant versions of quantum theory presents challenges, no comparable difficulties arise in the non-relativistic domain. Bohmian mechanics is naturally invariant under Galilean transformations. The GRW theory is invariant under orthochronous Galilean transformations, but not time-reversal invariant. It is instructive to examine how these theories satisfy Anderson’s criterion and the relativity principle for isolated subsystems.

\subsection{Bohmian mechanics}\label{ch2-bm}
Bohmian mechanics describes point-particles with positions ${\bf X}_k \in {\mathbb R}^{3}$, $k=1,\dots,N$, whose motion depends on the wave function $\psi(x)$, where $x=({\bf x}_1,\dots,{\bf x}_N) \in {\mathbb R}^{3N}$. The particle trajectories satisfy{\footnote{Throughout, units are assumed such that $\hbar=c=1$.}} 
\begin{equation}
\frac{d {\bf X}_k (t)}{dt}=  {\bf v}^{\psi_t}_k(X(t)), 
\label{ch2-ge}
\end{equation}
where $X(t) = ({\bf X}_1(t),\dots,{\bf X}_N(t))$ is the configuration, and
\begin{equation}
{\bf v}^\psi_k =  \frac{\hbar}{m_k} {\textrm{Im}} \frac{{\boldsymbol {\nabla}}_k \psi}{\psi}=  \frac{1}{m_k} {\boldsymbol {\nabla}}_k S , \qquad \psi = |\psi| \e^{\ii S/ \hbar}. 
\end{equation}
The wave function satisfies the usual Schr\"odinger equation
\begin{equation}
\ii \partial_t \psi_t(x) =  \left( -\sum^N_{k=1} \frac{\hbar^2 }{2m_k}\nabla^2_k + V(x) \right)  \psi_t(x) .
\label{ch2-se}
\end{equation}
This theory is  Galilean invariant (for a Galilean-invariant potential), since Galilean transformations map solutions $(X(t),\psi_t)$ of \eqref{ch2-ge} and \eqref{ch2-se} to solutions  $(X'(t),\psi'_t)$. For example, under a boost with velocity ${\bf v}$, we have 
\begin{align}
{\bf X}'_k(t) &= {\bf X}_k(t) + {\bf v} t  ,\\
\psi'_t(x)&= \mathcal{G}_{\bf v}(t) \psi_t(x) = \psi_t({\bf x}_1-  {\bf v} t,\dots, {\bf x}_N -  {\bf v} t) \exp\left[\ii \sum^N_{k=1} m_k \left( {\bf v}\cdot {\bf x}_k  - \frac{1}{2} v^2 t  \right) \right] ,  \end{align}
where
\be
\mathcal{G}_{\bf v}(t) = \ee^{-\ii Mv^2t/2}  \ee^{\ii M {\bf Q} \cdot {\bf v}}  \ee^{- \ii {\bf P} \cdot {\bf v}t} 
\en
is the boost operator, with $M=\sum^N_{k=1} m_k$ the total mass, and ${\bf Q} = \sum^N_{k=1} m_k {\bf q}_k/M$ and ${\bf P} = \sum^N_{k=1} {\bf p}_k$ respectively the center-of-mass position and momentum operators.

Since there are no absolute objects (in addition to those of Galilean space-time), the theory is Galilean invariant in Anderson's sense. The theory also satisfies the relativity principle. To see this, consider a system described by the coordinates ${\bf x}_1,\dots,{\bf x}_M$ and an environment described by the coordinates ${\bf x}_{M+1},\dots,{\bf x}_N$. Assume further that the potential $V(x)$ is separable, i.e.,  $V(x) = V_1({\bf x}_1,\dots,{\bf x}_M) + V_2({\bf x}_{M+1},\dots,{\bf x}_N)$. Then a wave function which is initially a product of the form
\be
\psi({\bf x}_1,\dots,{\bf x}_N) = \psi_s({\bf x}_1,\dots,{\bf x}_M) \psi_e({\bf x}_{M+1},\dots,{\bf x}_N) 
\label{ch2-sepnr}
\en
will remain a product at all times. As a consequence, the velocity field is   
\be
{\bf v}^{\psi_t}_k({\bf x}_1,\dots,{\bf x}_N) = 
\begin{cases}
	{\bf v}^{\psi_{s,t}}_k({\bf x}_1,\dots,{\bf x}_M)   & \text{if }\  k \leqslant M,\\
	  {\bf v}^{\psi_{e,t}}_k ({\bf x}_{M+1},\dots,{\bf x}_N) & \text{if } \ k > M.
\end{cases}
\en
Hence, it is clear that the relativity principle holds: a Galilean transformation can be applied to the system to obtain another solution of the dynamical equations. Of course, the assumption that the potential is separable can be relaxed. It suffices that the interaction between the system and its environment is negligible, for example when the interaction falls off with distance and the system is sufficiently far removed from its environment. (Furthermore, since the velocity of a configuration is only determined by the value of the wave function in the neighborhood of that configuration, it is actually sufficient that the wave function obtains the form \eqref{ch2-sepnr} in a sufficiently large region near the configuration. To make this more precise, the notion of conditional wave function could be employed, which is the natural definition of the wave function of a subsystem~\citep{duerr92a}.)

\subsection{GRW theory}\label{grw}
In the GRW theory \citep{bassi03}, the wave function undergoes collapses at random times. In between collapses, the wave function evolves according to the Schr\"odinger equation \eqref{ch2-se}. Under a collapse, the wave functions gets multiplied by a Gaussian function:
\be
\psi_t(x) \to \frac{\Lambda^{1/2}_I({\bf X}) \psi_t(x)}{\| \Lambda^{1/2}_I({\bf X}) \psi_t(x) \| }
\label{g1},
\en 
where 
\be
\Lambda_i({\bf x}) = \frac{1}{(2\pi \sigma^2)^{3/2}} \ee^{-\frac{({\bf q}_i - {\bf x})^2}{2\sigma^2}}
\en
is the collapse operator, with ${\bf q}_i$ the position operator for the $i$th particle. The label $I=1,\dots,N$ is random with uniform distribution and the collapse center ${\bf X}$ is random with distribution 
\be
P({\bf X} \in d^3y| \psi_t, I=i ) = \| \Lambda^{1/2}_{i}({\bf y}) \psi_t(x) \|^2 d^3 y.
\en
The time $T$ of the collapse is distributed according to the exponential distribution with rate $N\lambda$.

For the primitive ontology, two possibilities have been proposed \citep{allori11}: the mass density ontology and the flash ontology, resulting in the GRWm and GRWf theories. The mass density in the GRWm theory is 
\be
m({\bf x},t) = \sum^N_{k = 1} m_k \int d^3x_1 \dots d^3x_N \delta({\bf x}  - {\bf x}_k) |\psi_t({\bf x}_1,\dots,{\bf x}_N)|^2.
\label{g1.1}
\en
The flashes in the GRWf theory correspond to the (labeled) collapse events $({\bf X}_1,T_1,I_1)$,  $({\bf X}_2,T_2,I_2)$, \dots, where the ${\bf X}_i$ are the collapse centers, the $T_i$ the times of collapse, and $I_i$ labels the ``particle'' that undergoes collapse. Given an initial wave function $\psi_0$ at time $t_0$, the probability distribution for $n$ collapse events to have occurred in some time interval $[t_0,t]$, with $t_0< T_1< T_2 < \cdots < T_n < t $, is given by:
\begin{multline}
P^{\psi_0}({\bf X}_1 \in d^3 y_1, T_1 \in dt_1, I_1 = i_1, \dots, {\bf X}_n \in d^3 y_n, T_n \in dt_n, I_n = i_n)= \lam^n e^{-N\lam (t_n -t_0)} \\  
 \times \| \Delta_{i_n}^{1/2}({\bf y}_n) U_{t_n -t_{n-1}} \Delta_{i_{n-1}}^{1/2}({\bf y}_{n-1}) \cdots  \Delta_{i_{1}}^{1/2}({\bf y}_{1}) U_{t_1 -t_0} \psi_{t_0}\|^2 d^3 y_1dt_1 \dots d^3 y_ndt_n,
\label{g2}
\end{multline}
where $U_t = \exp(-\ii H t)$ it the time evolution operator corresponding to the Schr\"odinger equation \eqref{ch2-se}. This formula can actually be regarded as the defining law for the distribution of flashes, given an initial wave function $\psi_0$, without the need to consider the wave function collapses \eqref{g1}. It is this formulation that has been extended to the relativistic domain \citep{tumulka06a}.

The theory is invariant under orthochronous  Galilei transformations, but not under time reversal invariance. Because it is a stochastic theory, invariance requires not only that the dynamically allowed histories are mapped to dynamically allowed histories, but also that the probability distribution \eqref{g2} over these histories transforms properly. Consider, for example, the GRWf theory and a transformation $({\bf X}_i,T_i,I_i) \to ({\bf X}'_i,T'_i,I_i)$ and $\psi_0 \to \psi'_0$. If the distribution of $({\bf X}_i,T_i,I_i)$ is given by \eqref{g2}, then the distribution of transformed flashes is given by
\begin{multline}
P'({\bf X}'_1 \in d^3 x'_1, T'_1 \in dt'_1, I_1 = i_1, \dots, {\bf X}'_n \in d^3 x'_n, T'_n \in dt'_n,I_n = i_n) \\
= P^{\psi_0}({\bf X}_1 \in d^3 x_1, T_1 \in dt_1, I_1 = i_1,  \dots, {\bf X}_n \in d^3 x_n, T_n \in dt_n,I_n = i_n). 
\label{g4}
\end{multline}
The theory will be invariant under the transformation if this distribution is again of the form \eqref{g2}, i.e., it should be the case that $P' = P^{\psi'_0}$, so that 
\begin{multline}
P^{\psi'_0}({\bf X}'_1 \in d^3 x'_1, T'_1 \in dt'_1, I_1 = i_1, \dots, {\bf X}'_n \in d^3 x'_n, T'_n \in dt'_n,I_n = i_n) \\
= P^{\psi_0}({\bf X}_1 \in d^3 x_1, T_1 \in dt_1, I_1 = i_1,  \dots, {\bf X}_n \in d^3 x_n, T_n \in dt_n,I_n = i_n). 
\label{g5}
\end{multline}
This property is indeed satisfied for orthochronous Galilei transformations. Consider, for example, a Galilean boost:
\be
({\bf X}_i,T_i,I_i )\to  ({\bf X}_i + {\bf v} t,T_i,I_i ),
\en
\be
\psi_0 \to  \mathcal{G}_{\bf v}(t_0) \psi_0.
\en
The identity \eqref{g5} follows from the properties
\be
\mathcal{G}_{\bf v}(t) \Delta_{i}^{1/2}({\bf X})  =  \Delta_{i}^{1/2}({\bf X}- {\bf v} t) \mathcal{G}_{\bf v}(t) , \qquad \mathcal{G}_{\bf v}(t)U_{t -t'}  = U_{t -t'}\mathcal{G}_{\bf v}(t').
\en
 
As in the Bohmian case, this theory is invariant in Anderson's sense under orthochronous Galilean transformations. Also the relativity principle is satisfied. Namely, for a separable wave function of the form \eqref{ch2-sepnr}, we have that 
\begin{multline}
P^{\psi_0}({\bf X}_1 \in d^3 y_1, T_1 \in dt_1,  I_1 = i_1, \dots, {\bf X}_n \in d^3 y_n, T_n \in dt_n,I_n = i_n) =\\ P^{\psi_s}({\bf X}_{s_1} \in d^3 y_{s_1}, T_{s_1} \in dt_{s_1},  I_{s_1} = i_{s_1},  \dots, {\bf X}_{s_m} \in d^3 y_{s_m}, T_{s_m} \in dt_{s_m},I_{s_m} = i_{s_m}) \\ \times P^{\psi_e}({\bf X}_{e_1} \in d^3 y_{e_1}, T_{e_1} \in dt_{e_1}, I_{e_1} = i_{e_1},  \dots, \\ {\bf X}_{e_{n-m}} \in d^3 y_{e_{n-m}}, T_{e_{n-m}} \in dt_{e_{n-m}},I_{e_{n-m}} = i_{e_{n-m}}),
\end{multline}
where the $s_k \in \{1,\dots,n \}$ are such that $i_{s_k} \in \{1,\dots, M\}$ and $T_{s_1}<T_{s_2}<\dots$, i.e., they label the flashes of the system, and $e_k \in \{1,\dots,n \}$ are such that $i_{e_k} \in \{M+1,\dots, N\}$ and $T_{e_1}<T_{e_2}<\dots$, and label the flashes of the environment. As such, the system can be transformed, leaving the environment invariant, leading to a universal history that is dynamically allowed, with the correct probabilities. 

\subsection{Many-Worlds theory}\label{mwnr}
As mentioned in the introduction, according to the usual view of the Many-Worlds theory, the ontology is fully specified by the universal wave function, which evolves deterministically according to the Schr\"odinger equation \eqref{ch2-se}. There is no additional ontology in space-time. But there are also versions of the Many-Worlds theory that {\em do} take space-time and stuff in space-time as fundamental. One proposal is the Sm theory, which introduces the mass density, as defined in \eqref{g1.1} \citep{allori08}. This theory straightforwardly qualifies as substantively Galilean invariant according to both Anderson's criterion and the relativity principle. 

Galilean invariance follows from the fact that the mass density transforms as a scalar under Galilean transformations. For example, under a Galilean boost, where $\psi \to {\mathcal G}_{\bf v} \psi$, the mass density transforms as 
\be
m({\bf x},t) \to  m({\bf x} + {\bf v}t,t).
\en
In the absence of additional absolute objects, the theory is also Galilean invariant in Anderson's sense. 
To verify the relativity principle, consider a separable wave function of the form  \eqref{ch2-sepnr}. In that case, the mass density is 
\be
m({\bf x},t) = m_s({\bf x},t) +  m_e({\bf x},t),
\en
where $m_s({\bf x},t)$ and $m_e({\bf x},t)$ are respectively the mass densities determined by $\psi_s$ and $\psi_e$. Transforming just the system corresponds to $\psi_s \to \psi'_s$ and $m_s \to m'_s$. Since the transformed universal state is again dynamically allowed, the relativity principle is satisfied. (Actually, to apply the relativity principle, it is sufficient that the separable form obtains for a single world, where a ``world'' refers to a decohered branch of the wave function.)

\section{Lorentz-invariant Bohmian models}\label{ch2-rbm}
\subsection{Dynamics}
There have been various suggestions for Lorentz-invariant Bohmian models \citep{berndl96a,duerr99,horton01,dewdney02,goldstein03,horton04,nikolic05b,tumulka07,duerr14a,tumulka14,galvan15}. Here, we will consider a few representative examples in the simplified setting of a fixed number of free Dirac particles \citep{duerr99,duerr14a}. While there are no interactions, there can still be entanglement, which is the source of the non-locality.

\begin{figure}[t]
  \begin{center}
    \includegraphics[width=0.8\textwidth]{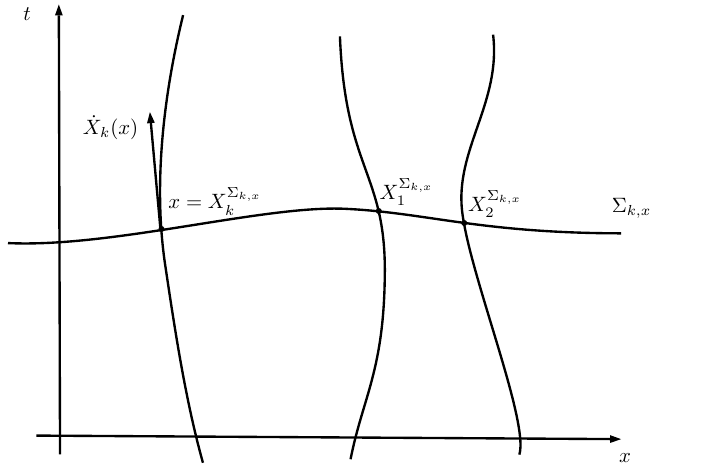}
\end{center}
\caption{The tangent vector $\dot X_k(x)$ of the $k$-th worldline at the point $x$ depends on the surface $\Sigma_{k,x}$ and the crossings of this surface with the other worldlines, cf.\ Eq.\ \eqref{ch2-guidance}. (Only one spatial dimension is shown.)}
\label{ch2-surface}
\end{figure}

In the multi-time picture, the wave function $\psi = \psi(x_1, ..., x_N)$ is a function on ${\mathcal M}^N$, with ${\mathcal M}$ Minkowski space-time, and takes values in the $N$-particle spin space $({\mathbb C}^4)^{\otimes N}$. To simplify the discussion that follows, we do not require the wave function to be completely anti-symmetric (although such a condition would be appropriate for a description of fermionic particles). The wave function satisfies the $N$ Dirac equations
\begin{equation}
\ii  \gamma^\mu_k \partial_{k,\mu} \psi  -  m_k \psi = 0,
\label{ch2-multitimedirac}
\end{equation}
with $k = 1, ..., N$.  Here, $\gamma^\mu_k = I \otimes \cdots \otimes I \otimes \gamma^\mu \otimes I \otimes \cdots \otimes I$ with the Dirac matrix $\gamma^\mu$ at the $k$-th of the $N$ places. There are also $N$ particles with worldlines $X_k(s)$, $k=1,\dots,N$, parameterized by $s$. In the models we will consider, the equations of motion for the particles are of the form
\be
{\dot X}_k(x) \propto  J^{\Si_{k,x}, \psi}_k\left(X^{\Si_{k,x}}_1,\dots,X^{\Si_{k,x}}_N\right).
\label{ch2-guidance}
\en
As illustrated in Fig.\ \ref{ch2-surface}, ${\dot X}_k(x)$ is the unit tangent vector to the $k$-th worldline at the  point $x \in {\mathcal M}$, $\Si_{k,x}$ is a space-like hypersurface (to be specified below) for the $k$th particle that contains $x$, and $X^{\Si_{k,x}}_i$ is the crossing of the $i$-th worldline with the surface $\Si_{k,x}$. The vector field $J^{\Si, \psi}_k$ is 
\be
\left[ J^{\Si, \psi}_k\right]^{\mu_k}(x_1,\dots,x_N) = J^{\mu_1 \dots \mu_N} (x_1,\dots,x_N) n_{\mu_1}(x_1) \dots {\widehat {n_{\mu_k}(x_k)}} \dots n_{\mu_N}(x_N),
\label{ch2-vel}
\en
where $n^\mu$ is the unit vector field normal to $\Si$,  
\be
J^{\mu_1 \ldots \mu_N} (x_1, \dots, x_N) = \overline{\psi} ( x_1, \ldots, x_N) \gamma_1^{\mu_1} \ldots \gamma_N^{\mu_N} \psi ( x_1 , \ldots, x_N),
\en
and the hat on the vector $n_{\mu_k}(x_k)$ indicates that it should be omitted from the expression.

\begin{figure}[t]
  \begin{center}
    \includegraphics[width=0.8\textwidth]{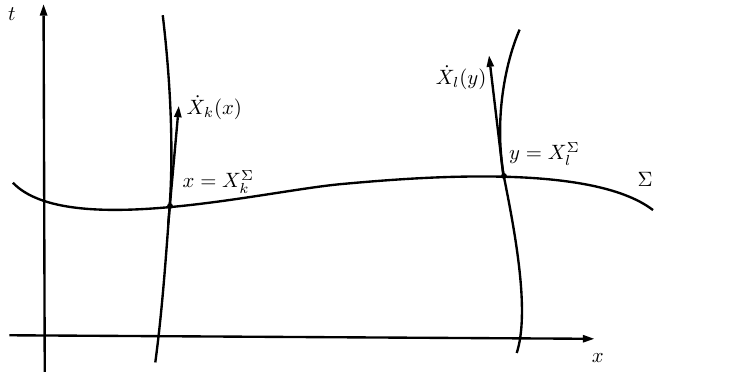}
\end{center}
\caption{Trajectories and tangent vectors in the case of a foliation, with a leaf $\Si$. In this case, $\Si_{k,x} = \Si_{l,y} = \Sigma$.}
\label{ch2-surface2}
\end{figure}

So, these models involve particle worldlines, the wave function, and the surfaces $\Si_{k,x}$. It remains to specify these surfaces. There are different possibilities. The surfaces may be determined by the wave function or the worldlines, or may satisfy some law, possibly also depending on the wave function or particle positions. The surfaces may or may not determine a foliation of space-time (see Fig.\ \ref{ch2-surface2}). 

A foliation ${\mathscr{F}}$ can be completely characterized by its unit normal vector field $n^\mu$. A unit vector field $n^\mu$ will determine a foliation iff $n \wedge \ud n = 0$, where $n =n_\mu \ud x^\mu$ is the one-form associated to the vector field. 

\begin{figure}[t]
  \begin{center}
    \includegraphics[width=0.8\textwidth]{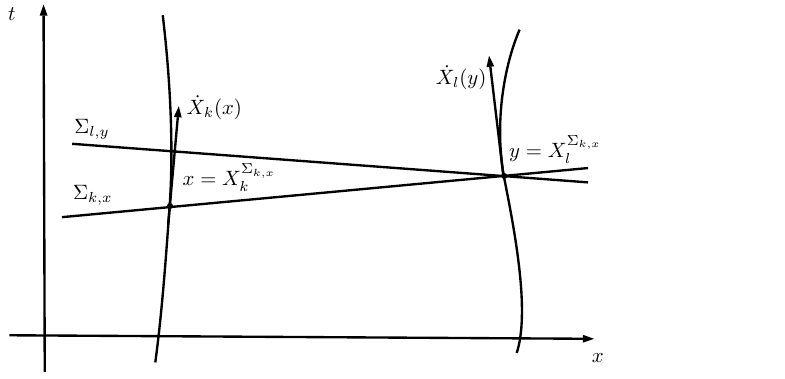}
\end{center}
\caption{Trajectories and tangent vectors in the case of the models \ref{ch2-5} and \ref{ch2-6}. The surfaces do not determine a foliation, since $\Sigma_{k,x} \neq \Sigma_{l,y}$, even though $y \in \Sigma_{k,x}$.}
\label{ch2-surface3}
\end{figure}

The following six choices for the surfaces $\Si_{k,x}$ are considered:
\begin{enumerate}[label=(\roman*)]
\item \label{ch2-1}
The $\Si_{k,x}$ are the leaves of a foliation ${\mathscr{F}}$, with unit normal vector field $n^\mu$ that satisfies \cite[p.\ 2736]{duerr99}:
\be
\pa_\mu n^\nu =0.
\label{ch2-l1}
\en
This law determines a foliation in terms of hyperplanes.{\footnote{This actually amounts to a Lorentz-invariant way of introducing a preferred Lorentz frame. The Newtonian dynamics can similarly be made Lorentz invariant by defining the temporal and spatial metrics as $n^\mu n^\nu$ and $n^\mu n^\nu - \eta^{\mu \nu}$ respectively, where $\eta^{\mu \nu}$ is the Minkowski metric.}} (Given a particular space-like hyperplane as an ``initial" leaf of the foliation, the law determines the whole foliation.)
\item \label{ch2-2}
The $\Si_{k,x}$ are the leaves of a foliation ${\mathscr{F}}$, with unit normal vector field $n^\mu$ that satisfies \citep{tumulka07}:
\be
\pa_\mu n_\nu - \pa_\nu n_\mu= 0.
\label{ch2-l2}
\en
This law implies that $n^\mu \pa_\mu n_\nu = 0$. Hence, the vector field $n^\mu$ does not change in the direction normal to the surfaces. (Again, given a particular space-like Cauchy{\footnote{A Cauchy surface crosses every inextendible time-like or light-like curve just once \citep{tumulka22}. A space-like hypersurface is not necessarily a Cauchy surface.}} hypersurface as an initial leaf of the foliation, the law determines the whole foliation.{\footnote{Specified as such, the foliation may have kinks, i.e., points where the leaves are not smooth. The Bohmian dynamics can still be defined for such foliations \citep{struyve14}, but such details do not concern us here.}} The distance between two surfaces, in the normal direction, will be constant along the surface.)
\item \label{ch2-3}
The $\Si_{k,x}$ are the leaves of a foliation ${\mathscr F}$, with unit normal vector field $n^\mu$ that satisfies \citep{galvan15}:
\be
n \wedge \ud n = 0.
\label{ch2-l3}
\en
This implies that there is a foliation which is not further restricted. (Given an initial space-like Cauchy surface, there are many foliations that include it as a leaf.) 
\item \label{ch2-4}
The $\Si_{k,x}$ are the leaves of a foliation that is covariantly determined by the wave function \citep{duerr99,duerr14a}. That is, the law specifies a map $\psi \to {\mathscr{F}}^\psi$, such that the diagram 
\begin{equation}
 \CD
\psi{}    @>>>    {\mathscr{F}}^\psi{}\\
@V U_\Lambda VV         @VV \Lambda V\\
\psi{}'  @>>>    {\mathscr{F}}^{\psi{}'} 
\endCD
\end{equation}
is commutative. $\Lambda$ is a Lorentz transformation, with ${\mathscr{F}}' = \Lambda {\mathscr{F}}^\psi{} $ the corresponding natural transformation of the foliation and $\psi{}' = U_\Lambda \psi$ is the corresponding transformation of the wave function, i.e., 
\be
\psi{}'(x_1,\dots,x_N)  = (U_\Lambda \psi)(x_1,\dots,x_N)= D_\Lambda \otimes \cdots \otimes D_\Lambda \psi(\Lambda^{-1} x_1,\dots,\Lambda^{-1} x_N),
\label{ch2-trans}
\en
where $D_\Lambda$ is the representative of $\Lambda$ in the representation of the Lorentz group for the Dirac theory \citep{schweber61}. 

An example is the map $\psi \to P^\mu$, with $P^\mu$ the total energy-momentum 4-vector determined by $\psi$ \citep{duerr14a}. This vector satisfies $\pa_\nu P^\mu = 0$, so that it determines a foliation in terms of hyperplanes normal to it.{\footnote{This foliation determines a center-of-mass frame for the wave function. As such, this proposal is akin to Bell's attempt to formulate a Lorentz-invariant version of Newtonian mechanics using the center-of-mass frame determined by the particles.}}

Explicitly, this vector is given by 
\be
P^\mu = \sum^N_{k=1} \int_\Sigma d \si_{\nu} (x) T^{\mu \nu}_k(x),
\en
where
\begin{multline}
T^{\mu_k \nu_k }_k(x_k) = \int_\Sigma d \si_{\nu_1}(x_1) \ldots {\widehat{\int_\Sigma d \si_{\nu_k}(x_k) }} \ldots \int_\Sigma d \si_{\nu_N}(x_N)  \\
\overline{\psi} ( x_1, \ldots, x_N) \gamma_1^{\nu_1} \ldots  \gamma_{k-1}^{\nu_{k-1}} \frac{\ii}{2} \left( \overleftrightarrow{\pa}^{\mu_k} \gamma_k^{\nu_k}+\gamma_k^{\mu_k} \overleftrightarrow{\pa}^{\nu_k} \right)  \gamma_{k+1}^{\nu_{k+1}}  \dots \gamma_N^{\nu_N} \psi ( x_1 , \ldots, x_N) ,
\label{emt}
\end{multline}
with $\Sigma$ some arbitrary space-like hypersurface and $\overleftrightarrow{\pa}^{\mu} = \frac{1}{2}(\overrightarrow{\pa}^{\mu} - \overleftarrow{\pa}^{\mu} )$.{\footnote{For a single particle, $T^{\mu \nu }(x)$ is the symmetrized energy-momentum tensor.}} This expression does not depend on the choice of $\Sigma$, since the divergence of the integrand with some $\pa_{\nu_i}$ gives zero \cite[p.\ 58]{schweber61}. This also implies that $T^{\mu_k \nu_k }_k(x_k)$ does not depend on the $x_i$, $i \neq k$. It follows similarly that $P^\mu$ is constant and does not depend on $\Sigma$.

\item \label{ch2-5}
The $\Si_{k,x}$ are the surfaces orthogonal to 
\be
P^\mu_k = \int_\Sigma d \si_{\nu} (x) T^{\mu \nu}_k(x).
\en
For each $k$, the constant $P^\mu_k$ determines a foliation in terms of hyperplanes. But this foliation might be different for different $k$.{\footnote{For a completely anti-symmetric wave function, all the $P^\mu_k$ are equal and the law reduces to the one considered in \ref{ch2-4}.}} So, it is not necessarily the case that the velocity fields are determined by a single foliation. It might be the case, as illustrated in Fig.\ \ref{ch2-surface3}, that $\Sigma_{k,x} \neq \Sigma_{l,y}$, where $y$ is the crossing of $\Si_{k,x}$ with the $l$-th worldline. 

\item \label{ch2-6}
The $\Si_{k,x}$ are the hyperplanes orthogonal to ${\dot X}_k(x)$. Generally, these surfaces do not determine a foliation (not even for a fixed $k$).

\end{enumerate}

Let us first mention some properties of these models. In each case, the single-particle law is
\be
{\dot X}^\mu (x) \propto J^\mu(x) = \overline{\psi} ( x) \gamma^{\mu} \psi ( x).
\label{ch2-single}
\en
Hence, the possible trajectories are the integral curves of the 4-current $J^\mu$, which does not depend on any surface.  

The dynamics is non-local, since for an entangled wave function the velocity of a particle will depend on the positions of the other particles along the relevant space-like hypersurface $\Sigma$. Such non-locality is unavoidable by Bell's theorem.

While there are influences over space-like distances, the trajectories are  never space-like. That is because the tangent vector ${\dot X}_k(x)$ is always time-like or light-like, i.e., $\left[ J^{\Si, \psi}_k\right]^{\mu_k}\left[ J^{\Si, \psi}_k\right]_{\mu_k} \geqslant 0$ for all $k$ and choices of space-like hypersurface $\Sigma$ \citep{duerr99}. (At least in the model \ref{ch2-1}, the particles typically do not reach the light-speed \citep{tausk10}.)

All these models are also Lorentz invariant (or, in Anderson's terminology, the Lorentz group forms the covariance group); Lorentz transformations map solutions of the dynamical equations to solutions. That is, a Lorentz transformation $\Lambda$ will map the worldlines $(X_1,\dots,X_N)$ to other dynamically allowed wordlines $(\Lambda X_1,\dots,\Lambda X_N)$, the wave function $\psi$ will be mapped to $U_\Lambda\psi$ (as defined in \eqref{ch2-trans}) and the foliation  ${\mathcal F}$ (if there is any) to $\Lambda{\mathcal F}$.

The models \ref{ch2-1}-\ref{ch2-3} contain extra space-time structure, namely the foliation. The trajectories depend on this foliation, but the foliation itself is independent of the particles. In the models \ref{ch2-4}-\ref{ch2-6}, there is no extra space-time structure. The surfaces $\Si_{k,x}$ are completely determined by the wave function in the models \ref{ch2-4} and \ref{ch2-5} and by the worldlines in model \ref{ch2-6}.

In the models \ref{ch2-1}-\ref{ch2-4}, the surfaces $\Si_{k,x}$ are part of a single foliation of space-time. For these models, there is a notion of quantum equilibrium, and it can be argued that these models reproduce the usual quantum predictions \citep{duerr99,lienert20,tumulka22}. Namely, in quantum equilibrium, the density of crossings of the trajectories with a leaf $\Si$ of the foliation with normal field $n^\mu$ is given by 
\be
\rho^{\Sigma,\psi} (x_1,\dots,x_N)  = J^{\mu_1 \dots \mu_N} (x_1,\dots,x_N) n_{\mu_1}(x_1) \dots  n_{\mu_N}(x_N),
\label{equi}
\en
relative to the volume defined by the induced metric on $\Sigma$, and evaluated on $\Sigma^N$. (In the case $\Si$ is a hyperplane, $\rho^{\Sigma,\psi}$ equals the familiar expression $\psi^\dagger \psi$, with $\psi$ the wave function on that hyperplane.) The crossings with a space-like Cauchy surface $\Si'$ that is not a leaf in the foliation will generally {\em not} be distributed according to $\rho^{\Si',\psi}$~\citep{berndl96a}. Nevertheless, it can be shown that actual measurements along an arbitrary space-like Cauchy surface will give results in accordance with the Born rule. This also means that the foliation itself is unobservable. Hence, while the trajectories depend on the foliation, the statistical predictions do not. (In quantum non-equilibrium, the foliation {\em would} be observable.)

In the models \ref{ch2-5} and \ref{ch2-6}, on the other hand, the surfaces (generally) do not belong to a single foliation.{\footnote{In \citep{horton01,dewdney02,goldstein03,horton04,tumulka14} other laws are considered that are not based on a foliation and which hence have the problem of statistical transparency. For example, in \citep{goldstein03,tumulka14}, the surfaces are given by the future light-cones and the contraction in \eqref{ch2-vel} is with the tangent vectors of the particles rather than the normals to the surface (which do not exist in this case).}} This has significant consequences. First, it is hard to establish the existence of solutions. This is especially the case for the model \ref{ch2-6}, where the stipulation of the surfaces is not independent of the particle dynamics \eqref{ch2-guidance}, potentially rendering the model overconstrained. Second, it also implies that it is hard to perform a statistical analysis. It is therefore not at all clear to what extent these models reproduce the usual quantum predictions. (In the non-relativistic limit, assuming that there is a frame where the velocities of all the particles are small compared to the light speed, the surfaces $\Si_{k,x}$ nearly agree, and presumably the predictions of non-relativistic quantum mechanics are recovered.)

\subsection{Anderson's criterion in terms of absolute objects}
According to Anderson's definition, the vector field $n^\mu$ and thus the corresponding foliation are absolute objects in the model~\ref{ch2-1}, but not in the models \ref{ch2-2} and \ref{ch2-3}. Namely, in the model \ref{ch2-1}, the space-time constancy of the vector field $n^\mu$ implies that any two allowed vector fields are related by a Lorentz transformation. This is not the case in the models \ref{ch2-2} and \ref{ch2-3}, because the law for $n^\mu$ allows for vector fields that are not connected by a Lorentz transformation. Hence, the group of Lorentz transformations is part of the symmetry group in Anderson's sense in the models \ref{ch2-2} and \ref{ch2-3}, but not in model~\ref{ch2-1}. (Actually, the full symmetry group is the Poincar\'e group, which includes space-time translations in addition to the Lorentz transformations.)

However, this conclusion seems counterintuitive. The concern parallels the one raised by Toretti in the context of Newtonian mechanics, cf.\ Section \ref{sli}. The foliation plays exactly the same structural role in all the models \ref{ch2-1}-\ref{ch2-3}. It therefore seems odd that the foliation is absolute in model~\ref{ch2-1}, but not in the models \ref{ch2-2} and \ref{ch2-3}. Moreover, if one wants to avoid the classification of  \ref{ch2-1} as being not Lorentz invariant according to Anderson's notion, then one could just extend the set of solutions by passing to the model \ref{ch2-2} or \ref{ch2-3}. If the actual world can be described by model \ref{ch2-1}, it can also be described by the models \ref{ch2-2} and \ref{ch2-3}. That the actual world involves a foliation that consists of hyperplanes, rather than curved surfaces, is then merely a contingency. 

Actually, Anderson introduced the concept of absolute objects to articulate the intuition that certain objects act without being acted back upon, thereby violating a generalized principle of action and reaction, see \citep[p.\ 339]{anderson67} and \citep[p.\ 169]{anderson71}. Yet in the present case, Anderson's definition fails to capture that idea. In each of the models \ref{ch2-1}-\ref{ch2-3}, the foliation acts on the particles without any back-reaction. However, as already observed by \citet{norton93}, the notion of action without back-reaction is difficult to formulate in a precise way. Here too, in the context of Bohmian mechanics, this idea seems immediately problematic. The wave function acts on the particles but without back-reaction, yet it does not seem to make sense to regard it as an absolute object.
 
The models \ref{ch2-4}-\ref{ch2-6} do not introduce extra space-time structure like a foliation; the ontology consists solely of the wave and the particles. As such, there are no absolute objects (other than the Minkowski metric). However, there still is a dynamically distinguished foliation in the case of \ref{ch2-4}. It is along the leaves of this foliation that the non-locality will be instantaneous and that the quantum equilibrium distribution has the form \eqref{equi}.

In summary, if Anderson’s criterion were adopted as a test for substantive Lorentz invariance, only model \ref{ch2-1} would fail to qualify. But this seems at odds with the intuition that the models \ref{ch2-1}-\ref{ch2-3} seem to stand on equal footing with respect to their claim to substantive Lorentz invariance.

\subsection{Relativity principle for subsystems}
To describe an isolated subsystem, consider a product wave function{\footnote{If the form \eqref{ch2-sep} holds on a surface $\Sigma_1 \times \cdots \times \Sigma_N$, with the $\Sigma_i$ Cauchy surfaces (e.g., $\Sigma_i$ constant-time surfaces in some Lorentz frame), then it will hold on ${\mathcal{M}}^N$. Note that this form cannot be obtained if the wave function is completely anti-symmetric. However, such a form could hold locally on ${\mathcal M}^N$, which would be sufficient to carry out a similar analysis. }} 
\be
\psi(x_1,\dots,x_N) = \psi_s(x_1,\dots,x_M) \psi_e(x_{M+1},\dots,x_N) .
\label{ch2-sep}
\en
The wave functions $\psi_s$ and $\psi_e$ will then separately satisfy the many-particle Dirac equations \eqref{ch2-multitimedirac}. Since  
\be
 J^{\Si, \psi}_k(x_1,\dots,x_N) = 
\begin{cases}
 J^{\Si, \psi_s}_k(x_1,\dots,x_M)  \rho^{\Sigma,\psi_e}( x_{M+1} , \ldots, x_N) & \text{if }\  k \leqslant M,\\
\rho^{\Sigma,\psi_s}( x_{1} , \ldots, x_M)  J^{\Si, \psi_e}_k(x_{M+1},\dots,x_N)   & \text{if } \ k > M,
\end{cases}
\en
the corresponding particle dynamics \eqref{ch2-guidance} can be written as
\be
{\dot X}_k(x) \propto 
\begin{cases}
 J^{\Si_{k,x}, \psi_s}_k(X^{\Si_{k,x}}_1,\dots,X^{\Si_{k,x}}_M)  & \text{if }\  k \leqslant M,\\
 J^{\Si_{k,x}, \psi_e}_k(X^{\Si_{k,x}}_{M+1},\dots,X^{\Si_{k,x}}_N) & \text{if } \ k > M. 
\end{cases}
\label{ch2-sepvel}
\en
For the system's particles $1,\dots,M$ to be isolated, it is also necessary that $J^{\Si_{k,x}, \psi_s}_k$ does not depend on the wave function $\psi_e$, nor on the particle positions $X_{M+1},\dots, X_N$. For the models considered here, there is only such dependence in the case \ref{ch2-4}, where the surface $\Si_{k,x}$ depends on the universal wave function $\psi$. (In model \ref{ch2-5}, the vectors $P^\mu_k$, $k=1,\dots,M$, and hence the corresponding surfaces $\Si_{k,x}$, only depend on $\psi_s$.) So, in the case of the model \ref{ch2-4}, there are no isolated subsystems, except in the special case where the wave function $\psi_s$ is completely separable, i.e.,
\be
\psi_s(x_1,\dots,x_M) = \psi_1(x_1) \dots \psi_M(x_M),
\label{ch2-sep2}
\en
because then the particles $1,\dots,M$ move independently of each other according to the single-particle dynamics \eqref{ch2-single}. If $\psi_s$ is {\em not} of that form, then the dynamics of (some of) the particles $1,\dots,M$ will depend on $\psi_e$. Namely, if $\psi_s$ is entangled, then the non-local influences between the particles $1,\dots,M$  will be instantaneous along the leaves of the foliation determined by the total wave function $\psi$. The upshot is that the relativity principle only holds for the special class of subsystems whose wave function is of the form~\eqref{ch2-sep2}. It does not hold for systems of arbitrary complexity (as characterized above) needed for a proper application of the principle.

So, except for the model \ref{ch2-4}, there exist isolated subsystems, and Lorentz transformations can be considered of these subsystems that leave the environment invariant. As is easy to see, applying such a transformation leads to a new solution only in the models \ref{ch2-2}, \ref{ch2-3}, \ref{ch2-5} and \ref{ch2-6}, but not in the model \ref{ch2-1}. Namely, in the model \ref{ch2-1}, a boost of the subsystem will not lead to a new solution, except again in the special case of a wave function $\psi_s$ which is completely separable. This is because the dynamics of the particles of the subsystem (with non-separable $\psi_s$) depend on a particular constant vector $n^\mu$. Hence, the dynamics of the transformed system will depend on the boosted vector $n'^\mu$. But the vector field $n^\mu$ should be constant over space-time and cannot vary locally. No such problem arises in the models \ref{ch2-2} and \ref{ch2-3}. In those models, a transformation of a subsystem includes not only a transformation of the particle variables but also of the vector field $n^\mu$ in the region occupied by the subsystem. In that case, a local change of $n^\mu$ still results in a dynamically allowed vector field.

In summary, the models \ref{ch2-2}, \ref{ch2-3}, \ref{ch2-5} and \ref{ch2-6} satisfy the relativity principle for isolated subsystems. The difference with Anderson's approach is that now also model~\ref{ch2-4} fails to qualify as  substantively Lorentz invariant. As with Anderson's approach, it seems counterintuitive that model \ref{ch2-1} does not  qualify as  substantively Lorentz invariant, while models \ref{ch2-2} and \ref{ch2-3}, which enlarge the set of solutions  of model \ref{ch2-1}, do. (One could actually enlarge the solution space of \ref{ch2-4} too, to obtain a model that satisfies the relativity principle, by stipulating that in the case of a separable wave function \eqref{ch2-sep}, the particle dynamics for the subsystems depends on surfaces that are determined by the energy-momentum 4-vector either corresponding to $\psi_s$ or $\psi$. But such a move does not seem amenable to the case where interactions are included. In that case, the separability of the universal wave function into subsystem and environment wave functions is never exact, and hence it would not be clear how to formulate the universal law.) 

The findings are summarized in Table \ref{tablesum}.

\begin{table}[t]
\begin{center}\begin{tabular}{|c c c c c c c c c|} 
 \hline
Model &\ref{ch2-1} & \ref{ch2-2} & \ref{ch2-3} & \ref{ch2-4} & \ref{ch2-5} & \ref{ch2-6} & rGRW & Sm\\ [0.5ex] 
\hline\hline
 Anderson's criterion & \ding{55} & \checkmark &  \checkmark & \checkmark & \checkmark &  \checkmark & \checkmark & \checkmark \\ 
 \hline
 Relativity principle &  \ding{55} & \checkmark &  \checkmark & \ding{55} & \checkmark &  \checkmark & \checkmark  & \checkmark \\
 \hline
\end{tabular}\end{center}
\caption{The appraisals of the Bohmian models \ref{ch2-1}-\ref{ch2-6}, the rGRWf model and the Sm theory, delivered by two possible characterizations of substantive Lorentz invariance: Anderson's criterion based on absolute objects and the relativity principle for subsystems. The symbol \checkmark indicates that the model is substantively Lorentz invariant, while  \ding{55} indicates that it is not.}
\label{tablesum}
\end{table}

\section{Lorentz-invariant spontaneous collapse models}\label{ch2-sct}
Among the relativistic spontaneous collapse models \citep{tumulka06a,tumulka21,pearle15}, we focus on the rGRWf theory \citep{tumulka06a,tumulka22}, since it is formulated for non-interacting Dirac particles, like the Bohmian theories discussed above. This model is invariant under orthochronous Lorentz transformations. It will be shown that both criteria for substantive Lorentz invariance are satisfied.

The model is a generalization of the GRWf model considered in Section \ref{grw}, based on the flash ontology.{\footnote{The mass density ontology of the GRWm theory can also be extended to the relativistic domain~\citep{bedingham14}.}} The flashes are denoted by $X^i_k$, where $k=1,\dots,N$ is the particle label and $i=1,\dots,n_k$ labels the subsequent flashes for the $k$-th particle.{\footnote{This notation is different from the one used in section \ref{grw}, but more convenient in the present case, because of the absence of interactions.}} The probability distribution of these flashes is determined by a wave function $\psi$ satisfying the multi-time Dirac equations \eqref{ch2-multitimedirac}. To each generation of flashes, a new ``collapsed'' wave function can be associated, but as in the GRWf case, these need not be considered to specify the law for the flashes. For a wave function $\psi$,
the joint probability distribution for the flashes $X^i_k$ to occur in the volume elements $d^4x^i_k$ is
\begin{multline}
P^\psi(X^i_k \in d^4x^i_k; k=1,\dots,N; i=1,\dots,n_k)\\
= \left\langle \psi_0 \left| E^{(n_1)}\left(\prod_{i=1}^{n_1}d^4x^i_1\right) \otimes \cdots \otimes E^{(n_N)}\left(\prod_{i=1}^{n_N}d^4x^i_N\right)\right| \psi_0\right\rangle,\label{ch2-prob}
\end{multline}
where $\psi_0$ is the restriction of $\psi$ to some surface $\Si_1 \times \cdots \times \Si_N$, with $\Si_i$, $i=1,\dots,N$ arbitrary space-like Cauchy surfaces. The $E^{(n_k)}$ are operators acting only on the $k$-th particle subspace \cite[p.\ 836]{tumulka06a}. Their explicit expression does not concern us here. For our purposes it suffices to note the following two properties. 

First, for an orthochronous Lorentz transformation $X^i_k \to X'^i_k=\Lambda X^i_k$, $\psi \to \psi' = U_\Lambda \psi$, $d^4x^i_k \to d^4x'^i_k = d^4\Lambda x^i_k$, the joint probability distribution satisfies
\begin{multline}
P^{\psi'}(X'^i_k \in d^4x'^i_k; k=1,\dots,N; i=1,\dots,n_k) = \\
P^\psi(X^i_k \in d^4x^i_k; k=1,\dots,N; i=1,\dots,n_k).
\label{ch2-liflash}
\end{multline}
This implies that the model is invariant under these transformations of the whole universe. Second, if the wave function is separable, i.e., of the form \eqref{ch2-sep}, then because of the form \eqref{ch2-prob} of the probability distribution, 
\begin{multline}
P^\psi(X^i_k \in d^4x^i_k; k=1,\dots,N; i=1,\dots,n_k) = \\
P^{\psi_{s}}(X^i_k \in d^4x^i_k; k=1,\dots,M; i=1,\dots,n_k)\\
\cdot P^{\psi_{e}}(X^i_k \in d^4x^i_k; k=M+1,\dots,N; i=1,\dots,n_k).
\end{multline}
Together with the first property and applying a similar argument as for Lorentz transformations of the whole universe, it is clear that the relativity principle for subsystems is satisfied.

Furthermore, the model has no absolute objects (beyond the Minkowski metric) and so the symmetry group includes the orthochronous Lorentz group according to Anderson's  criterion.

\section{Many-Worlds theory}\label{ch2-mw}
The Sm version of the Many-Worlds theory considered in Section \ref{mwnr} extends straightforwardly to the relativistic domain \citep{allori08}. In this setting, the relativistic generalization of the mass density is given by the energy-momentum tensor
\be
{\mathcal M}^{\mu \nu}(x) = \langle \psi | {\mathcal T}^{\mu \nu}(x) | \psi \rangle,
\en
where ${\mathcal T}^{\mu \nu}(x)$ is the energy-momentum tensor operator \citep{allori08}. For the non-interacting Dirac theory, this quantity equals ${\mathcal M}^{\mu \nu}(x) = \sum_k T^{\mu \nu}_k(x)$, where $T^{\mu \nu}_k(x)$ is given in Eq.~\eqref{emt}. As a tensor field, ${\mathcal M}^{\mu \nu}(x)$ possesses the appropriate transformation properties under Lorentz transformations. As in the non-relativistic case, it is straightforward to see that, according to both Anderson’s principle and the relativity principle, the theory qualifies as substantively Lorentz invariant.

\section{Conclusion}\label{ch2-con}
We have examined two possible ways of characterizing what substantive Lorentz invariance might amount to and found that most of the models considered here satisfy both characterizations. At the same time, some Bohmian models that do not appear to be substantively Lorentz invariant are nevertheless not ruled out by these criteria, suggesting that a more stringent criterion may be required.

In any case, Lorentz invariance is only one among several features one might reasonably demand of a fundamentally relativistic theory \citep{duerr14a,allori22,tumulka22}. Another natural requirement is locality. However, this requirement conflicts with Bell's theorem, which implies that any empirically adequate theory must involve non-local influences. A possible response is to avoid such influences by adopting an ontology that is itself non-local, that is, an ontology that is not restricted to variables that are well localized in space-time (called local beables by \citet{bell76a}), but also includes non-local variables. However, allowing for non-local variables just as well appears to conflict with relativity.

Further criteria may also be considered. For example, in the context of experimental tests, the locality condition in Bell's theorem is often analyzed as the conjunction of parameter independence and outcome independence. It has frequently been argued that Bohmian theories are less compatible with relativity than spontaneous collapse theories because they violate parameter independence \citep{jarrett84,shimony84,butterfield92,ghirardi96,ghirardi10,myrvold16,tumulka22}. Conversely, as argued by \citet{allori22}, spontaneous collapse theories may be regarded as less relativistic than Bohmian theories because they exhibit non-locality even at the level of a single particle, unlike Bohmian theories, where non-locality arises only in entangled multi-particle systems.

These criteria---both those discussed here and others found in the literature---should perhaps not be understood as strict requirements for a relativistic quantum theory. Rather, they may serve to highlight which features, in addition to locality, might have to be abandoned in the transition from classical to quantum theory. Since spontaneous collapse models and no-collapse theories such as Bohmian mechanics make empirically distinct predictions, this is not merely a metaphysical issue, but also an empirical one. It may ultimately turn out that Lorentz invariance holds only at the statistical level, rather than at the fundamental level, as in Bohmian theories with a preferred foliation of space-time. This would not imply a return to pre-relativistic space-times, but rather a modification of the relativistic space-time structure required to accommodate the non-locality~\citep{maudlin19b}.

\section{Acknowledgments}
This work is supported by the Research Foundation Flanders (Fonds Wetenschappelijk Onderzoek, FWO), Grant No.\ G0C3322N. It is a pleasure to thank David Albert, Detlef D\"urr, Sheldon Goldstein, Travis Norsen, Roderich Tumulka and Nino Zangh\`i for useful discussions over the years, as well as Guido Bacciagaluppi, Fred Muller and Sylvia Wenmackers for useful comments on the manuscript.

\bibliographystyle{apalike}
 \bibliography{ref.bib}

\end{document}